\documentclass[fleqn,twoside]{article}
\usepackage{espcrc2}

\usepackage{graphicx,amssymb,latexsym,amsmath}
\usepackage[figuresright]{rotating}

\title{The deconfining phase transition in SU($N_c$) gauge
theories\thanks{Talk by U.~Wenger.}}

\author{B. Lucini\address[Oxf]{Theoretical Physics, 
        Oxford University, \\ 
        1 Keble Road, Oxford OX1 3NP, United Kingdom}%
        \thanks{EU Marie Curie fellow.},
        M. Teper\addressmark[Oxf],
        U. Wenger\addressmark[Oxf]\thanks{PPARC SPG fellow.}}
       
\begin{document}

\begin{abstract}
We report on our ongoing investigation of the deconfining phase
transition in SU(4) and SU(6) gauge theories. We calculate the
critical couplings while taking care to avoid the influence of a
nearby bulk phase transition.  We determine the latent heat of the
phase transition and investigate the order and the strength of the
transition at large $N_c$.  We also report on our determination of the
critical temperature expressed in units of the string tension in the
large $N_c$ limit.
\vspace{1pc}
\end{abstract}

\maketitle

\section{INTRODUCTION}
The large-$N_c$ physics of QCD and SU($N_c$) gauge theories is of
great theoretical and phenomenological interest. Lattice calculations
of string tensions, glueball masses etc.~\cite{blmt-glue} have
confirmed that a smooth large-$N_c$ limit is achieved by keeping fixed
the 't Hooft coupling $g^2 N_c$ and that, as expected, the leading
corrections are $O(1/N_c^2)$.

An interesting question concerns the order of the deconfining phase
transition at $N_c \rightarrow \infty$. It has been argued
\cite{PisarskiTytgat} that the transition in SU(3) may be accidently
first order, due to a cubic term in the effective potential, but in
general is second order in SU($N_c$). Our ongoing lattice study
\cite{blmtuw:2002}, on which we report here, is addressing this
question and thereby improving on earlier results
\cite{WingateOhta,Gavai}.

We use the standard plaquette action on $L^3 \times L_t$ lattices with
periodic boundary conditions.  The deconfining phase transition is
studied by keeping $L_t$ fixed and varying $\beta = 2 N_c/g^2$ so as
to pass the temperature $T = 1 / a(\beta) L_t$ through the transition.

In order to distinguish the confining and the deconfining phases we
define the spatial average $\bar l_p$ of the Polyakov loop and
correspondingly the spatial average $\bar u_p$ of the pla\-quet\-te
variable $\text{Tr} U_p$.  The two order parameters so defined can be
used to characterise the order of the phase transition. For a first
order transition, e.g., we should find tunneling between the phases
near the critical temperature $T_c$ and subsequently a double peak
structure in the probability distributions of $\bar l_p$ and $\bar
u_p$.

In order to determine the critical couplings $\beta_c(V)$ at finite
volume $V=L^3$ we define a normalised Polyakov loop susceptibility
\begin{equation*}
\frac{\chi_l}{V}  = 
\langle |{\bar l}_p|^2 \rangle - {\langle |{\bar l}_p| \rangle}^2
\end{equation*}
and, for each volume $V$, obtain $\chi_l$ as a continuous function of
$\beta$ using standard reweighting techniques \cite{reweight}. Then we
determine $\beta_c$ from the location at which the susceptibility has
its maximum.

In complete analogy we can define the specific heat $C(\beta)$ from
the average plaquette $\bar u_p$,
\begin{equation*}
\frac{1}{\beta^2} C(\beta) 
= 
\frac{\partial}{\partial\beta} 
\langle \bar{u}_p  \rangle
=
{N_p}
\langle \bar{u}_p^2 \rangle 
- 
{N_p}
{\langle \bar{u}_p  \rangle }^2,
\end{equation*}
where $N_p=6L^3L_t$ is the total number of pla\-quet\-tes. The value
of $\beta$ where $C$ has its maximum provides a different definition
for a critical coupling which however should agree with the one from
$\chi_l$ in the thermodynamic limit. Moreover we can use the finite
size scaling behaviour of the peak value $C(\beta_c,V)$ as a criterion
for the determination of the order of the transition. To be more
precise, we expect the leading finite size scaling
\begin{equation*}
\frac{C(\beta_c, V)}{V} 
= 
\begin{cases}
a_1 L^{\rho - d} + a_2 L^{-d}, & \text{2nd order},\\
a_1 + a_2 L^{-d}, & \text{1st order},
\end{cases}
\end{equation*}
where $\rho < 3$ is some combination of critical exponents and $d=3$. 

An important point we need to address is the influence of a bulk phase
transition which can be close to the physical deconfining
transition. Our choices of $L_t$ for the different gauge groups ensure
that the two transitions are well separated and can unambiguously be
distinguished.

\section{$T_c$ IN SU(4)}
In SU(4) we perform a finite size scaling study at $L_t = 5$ on $L =
12, 14, 16, 18, 20$ lattices with between 200 and 300 ksweeps at each
$\beta$-value and lattice size.  For $\beta$-values close to $\beta_c$
we find clear tunneling transitions both in $\bar l_p$ and $\bar
u_p$. The probability distributions of these quantities exhibit clear
double peaking which is characteristic for a first order transition.

\begin{figure}[tb]
\includegraphics[angle=-90,width=7.5cm]{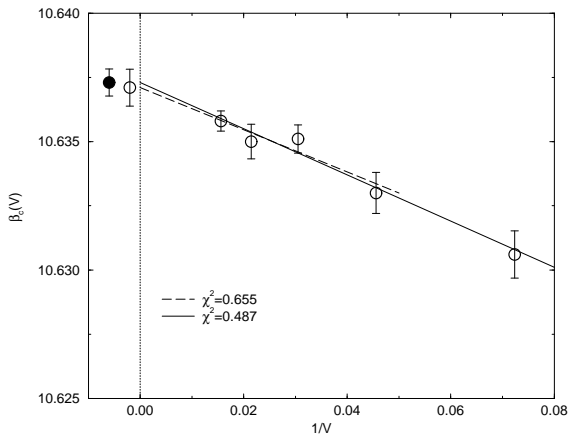}
\vspace{-1.0cm}
\caption{Finite size scaling of $\beta_c(V)$ for $L_t=5$ in SU(4).}
\label{fig:FSS_beta_c_from_L}
\end{figure}

In fig.~\ref{fig:FSS_beta_c_from_L} we plot our values of $\beta_c(V)$
against the inverse spatial volume expressed in units of the
temperature $T$ together with the extrapolations to infinite volume.
We note that if the transition is first order, as ours clearly is,
then the leading finite-$V$ correction should be
\begin{equation}\label{eq:FSS_beta_c}
\beta_c(V) =  \beta_c(\infty) - \frac{h}{VT^3},
\end{equation}
where $h$ is independent of the lattice action. We find that all our
values of $\beta_c$ are consistent with eq.~(\ref{eq:FSS_beta_c}) and
we obtain $\beta_c(\infty) = 10.63709(72)$ in the $V \rightarrow
\infty$ limit. Moreover the value of $h = 0.09\pm 0.02$ we thus
extract is close to the SU(3) value, $h \simeq 0.1$ \cite{Tsu3}. This
suggests that $h$ depends only weakly on $N_c$ and we will therefore
use this value to extrapolate to $V=\infty$ in the SU(6) case where we
do not perform an explicit finite volume study.

From $\beta_c(\infty)$ we obtain the critical temperature $T_c$ in
terms of the string tension $\sigma$ \cite{blmt-glue},
\begin{equation}\label{eq:C7}
\frac{T_c}{\sqrt{\sigma}}
=
0.6024 \pm 0.0045 \quad \text{at} \, \, a=1/5T_c.
\end{equation}
Our preliminary $L_t=6$ calculations give a value
$\beta_c(\infty)=10.780(10)$ which translates to $T_c/\sqrt{\sigma} =
0.597(8)$. This, taken together with the value in eq.~(\ref{eq:C7}),
yields the extrapolated continuum value
\begin{equation}\label{eq:C9}
\lim_{a\to 0}\frac{T_c}{\sqrt{\sigma}}
=
0.584 \pm 0.030 \quad \text{in SU(4)}.
\end{equation}

\begin{figure}[t]
\includegraphics[angle=-90,width=7.5cm]{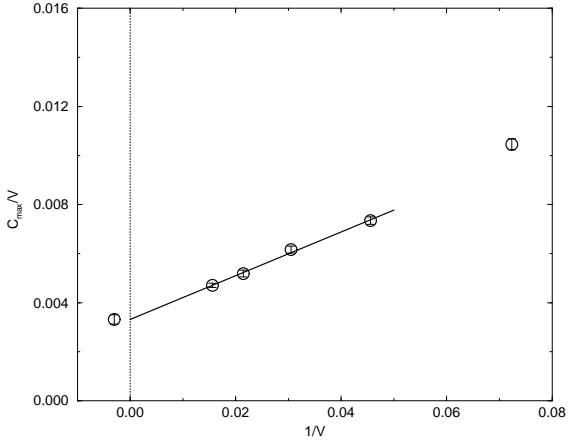}
\vspace{-1.0cm}
\caption{Normalised specific heat $C(\beta_c,V)/V$ in SU(4) at
$L_t=5$ plotted against $1/V$.}
\label{fig:cmax_FSS}
\end{figure}

In fig.~\ref{fig:cmax_FSS} we plot $C(\beta_c)/V$ against $1/V$ and
note that the scaling is certainly consistent with a first order one.
The intercept at $1/V = 0$ provides a measure of the latent heat,
$\Delta e$, since
\begin{equation*}
\lim_{V\to\infty} 
\frac{C(\beta_c,V)}{\beta_c^2 V} 
\propto
(\langle \bar{u}_{p,c} \rangle - \langle\bar{u}_{p,d}\rangle)^2
=
\Delta e^2,
\end{equation*}
where $\bar{u}_{p,c}$ and $\bar{u}_{p,d}$ are the average pla\-quette
values at $\beta_c$ in the confined and deconfined phases,
respectively. We obtain $\Delta e(N_c=4) = 0.00197(5)$ and observe
that if we take the $L_t=4$ SU(3) latent heat in \cite{ABS} and
naively scale it to $L_t=5$, we obtain $\Delta e(N_c=3) \simeq 0.0013$
which is substantially smaller than our above SU(4) value. This shows
explicitly that the SU(4) transition is more strongly first order than
the SU(3) one.

\section{$T_c$ IN SU(6)}
In SU(6) we perform a study on a $16^3 \times 6$ lattice at values of
$\beta$ close to $\beta=24.845$. In physical units this is like a
$14^3$ lattice with $L_t=5$ and thus we are confident from our SU(4)
calculations to be close to the thermodynamic limit.

As before we observe well defined tunneling between confined and
deconfined phases, characteristic of a first order phase transition.
Using the reweighting technique we extract a critical value
$\beta_c=24.850(3)$ which we extrapolate to $V=\infty$ using
eq.~(\ref{eq:FSS_beta_c}) with the corresponding value of $h = 0.09$
from SU(4). We obtain $\beta_c(\infty) = 24.855\pm0.003$ which
translates into
\begin{equation}\label{eq:D1}
\frac{T_c}{\sqrt{\sigma}}
=
0.588 \pm 0.002 \quad \text{at} \, \, a=1/6T_c
\end{equation}
using the values of the SU(6) string tension in \cite{Pisa1}.
Preliminary $L_t=5$ calculations provide an estimate $\beta_c(\infty)
= 24.519(33)$ which translates to $T_c/\sqrt{\sigma} =
0.580(12)$. Taken together with the value in eq.~(\ref{eq:D1}) this
gives the continuum value
\begin{equation}\label{eq:D2}
\lim_{a\to 0}\frac{T_c}{\sqrt{\sigma}}
=
0.605 \pm 0.026 \quad \text{in SU(6)}.
\end{equation}

As for the strength of the first order transition, we note that an
estimate of the latent heat, $\Delta e$, in SU(6) is larger than the
one on a directly comparable $16^3 6$ SU(4) calculation indicating
that the transition is certainly not weakening as $N_c \rightarrow
\infty$.

\section{CONCLUSIONS}
We find that SU(4) gauge theory at $L_t = 5$ reveals a first order
deconfining phase transition with a latent heat that is not
particularly small. We also find evidence that the SU(6) transition at
$L_t = 6$ is first order as well and certainly not weaker than the one
in SU(4). We come to that conclusion by comparing the appropriately
rescaled latent heat expressed in terms of $T_c$ for the different
gauge groups.  In doing so we take care to avoid confusion with a
nearby bulk phase transition which, on our lattices, can unambiguously
be identified.

The $N_c$-dependence of the critical temperature when expressed in
units of the string tension, $T_c/\sqrt{\sigma}$, appears to be weak.
From a fit we obtain
\begin{equation}\label{E1}
\frac{T_c}{\sqrt{\sigma}}
=
0.582(15) + \frac{0.43(13)}{N_c^2}.
\end{equation}

Finally, we conclude that the SU($N_c$) transition at
\mbox{$N_c=\infty$} is first order and not particularly weak.


\end{document}